  [2004/06/22 1.2 (KOF); 2001/04/25 1.1 (PWD)]

\documentclass[a4paper,twocolumn]{Gaia2004} % European paper size
\usepackage{times}      % for font
\usepackage{epsfig}     % for figure inclusion
\usepackage{natbib}     % for bibliography
\title{Variable stars as standard candles and stellar tracers}

\author{G. Bono$^1$, ~M. Cignoni$^2$}
\affil{INAF - Osservatorio Astronomico di Roma, Via Frascati 33, 00040 Monte 
Porzio Catone, Italy; bono@mporzio.astro.it\\
Dipartimento di Fisica "E. Fermi", Univ. di Pisa, Largo B. Pontecorvo 2, 
56127 Pisa, Italy; cignoni@df.unipi.it} 

\bibpunct{(}{)}{;}{a}{}{,}  % to set bibliography punctuation to A&A style

\begin{document}


\maketitle

\begin{abstract}
We review the current status of classical Cepheid and RR Lyrae distance scales. 
We discuss the most recent theoretical and empirical findings together with 
pros and cons of different methods. The use of Cepheids and RR Lyrae stars as 
stellar tracers are also discussed. Preliminary findings concerning the 
impact that GAIA will have on the calibration of distance determinations 
and astrophysical parameters are also presented. 
\end{abstract}

%%%%%%%%%%%%%%%%%%%%%%%%%%%%%%%%%%%%%%%%%%%%%%%%%%%%%%%%%%%%%%%%%%%%%%%%%%%%%%%%%
\section{Introduction}

During the last few years, the field of variable stars experienced 
a vivid interest from the astronomical community. The reasons are manifold. 
The use of the near-infrared interferometer (VINCI) available at the ESO 
Very Large Telescope has provided the angular diameter measurement for seven 
classical Cepheids (Kervella et al. 2004a). The new data, together with data already 
available in the literature, yield new calibrations  of the Period-Radius and 
Period-Luminosity (PL) relation. In particular, a very good agreement it has been 
found between interferometric and the infrared 
surface brightness version of the Baade-Wesselink (BW) method (Kervella et al. 2004b).    
However, the key issue concerning Cepheid distances is whether the PL relation 
is Universal. This is a classical problem which was addressed in two pioneering 
investigations by Kraft (1963) and Gascoigne (1969). Recent predictions based 
on pulsation models suggested either a marginal dependence on chemical composition 
(Baraffe, \& Alibert 2001), or a systematic shift toward lower effective temperatures 
when moving from metal-poor to metal-rich Cepheids (Bono et al. 1999a,b). 
In particular, the latter set of models predicted that metal-poor Cepheids are, 
at fixed period, brighter than metal-rich ones. This finding was at variance with 
empirical estimates. However, new empirical evidence appear to support theoretical 
predictions (Tammann et al. 2003; Sandage et al. 2004; Kanbur \& Ngeow 2004), but  
the problem is far from being settled, and indeed Storm et al. (2004) using 
a good sample of Galactic and Small Magellanic Cloud Cepheids found an opposite 
trend.  

These observational results are based on new photometric and radial 
velocity measurements. New high resolution and high signal to noise spectra 
to estimate heavy element abundances in classical Cepheids are quite limited. 
Recent detailed investigations for Galactic Cepheids 
have been provided by Fry \& Carney (1997) and by Andrievsky et al. (2004). 
However, accurate metal abundances are only available for two dozen 
Magellanic Cepheids (Luck et al. 1998). Note that Magellanic Cepheids play 
a crucial role in the metallicity dependence. They are systematically 
metal-poor relative to Galactic Cepheids and located at the same 
distance. Recent high resolution spectra collected with UVES available 
at VLT are filling this gap (Romaniello et al. 2004). The spectroscopic 
approach appears very promising, and indeed new spectra and homogeneous
distances for three dozen Galactic and Magellanic Cepheids indicate 
that V-band PL relation does depend on the metal abundance at 99.6\% 
confidence level. Moreover, these data suggest that steady linear decrease 
in Cepheid luminosity -when moving from metal-poor to metal-rich Cepheids- 
can be excluded at 99,95\% confidence level (Romaniello et al. 2004).        

Theoretical and empirical investigations for RR Lyrae stars produced both 
good and bad news. Recent findings concerning the near-infrared ($J,K$) 
PL relation of RR Lyrae stars suggest that distances based on this method
are marginally affected by systematic uncertainties due to evolutionary 
effects, and reddening corrections. Moreover, synthetic HB diagrams 
(Catelan et al. 2004; Cassisi et al. 2004) indicate that these relations 
depend marginally on metal abundance in the metal-poor regime ($Z\le 0.006$).  
Hence, in the near future, we should be able to constrain the absolute distance 
to Galactic Globular clusters (GGCs) with an accuracy better than 5\% 
(Gratton et al. 2003). However, it is worth mentioning that 
such an accuracy implies an uncertainty in absolute magnitude of the 
order of 0.1 mag. This means that we cannot use RR Lyrae stars to 
constrain the input physics currently adopted to construct HB models.    
In fact, the typical difference is of the same order of magnitude 
(see Fig. 13 in Pietrinferni et al. 2004). Note that trigonometric 
parallaxes for field RR Lyrae measured by HIPPARCOS (Fernley et al. 1998)
and for RR Lyr itself measured by Benedict et al. (2002) using the FGS on 
board of HST did not solve this problem (Bono et al. 2002).  

Although RR Lyrae stars have been subject of enormous observational efforts, 
we still lack several fundamental information on these objects. The BW technique 
has been widely applied to RR Lyrae stars (Cacciari et al. 2000), but an empirical 
PR relation has not yet been provided (Marconi et al. 2004). The same outcome 
applies to chemical abundances. High resolution spectra for field RR Lyrae 
have been collected by Clementini et al. (1995), Fernley \& Barnes (1997), 
and by Solano et al. (1997). More recently low-resolution spectra for a 
sizable sample ($\sim 100$) of Large Magellanic Cloud RR Lyrae stars have 
been collected by Gratton et al. (2004). However, we still lack a 
detailed analysis of heavy element abundances in cluster RR Lyrae
variables.  

GAIA will play a fundamental role in these open problems, because 
it will supply homogeneous multiband 
photometric data, radial velocities, chemical compositions, and 
geometrical distances for large samples of radial variables with 
an unprecedented accuracy. In the following, we discuss the use of 
Classical Cepheids (section 2) and RR Lyrae (section 3) stars as 
standard candles and stellar tracers. The topics discussed in the 
paper shall be considered as provisions for the journey we have 
undertaken while waiting for the GAIA database.  

%%%%%%%%%%%%%%%%%%%%%%%%%%%%%%%%%%%%%%%%%%%%%%%%%%%%%%%%%%%%%%%%%
\section{Classical Cepheids}

Current theoretical predictions for the PL and the PLC relations 
of classical Cepheids rely on a fundamental interplay between evolutionary 
and pulsation predictions. Pulsation models are constructed by assuming 
a Mass-Luminosity (ML) relation predicted by evolutionary models 
(Bertelli et al. 1993; Bono et al. 1999a,b; Alibert et al. 1999; 
Baraffe \& Alibert 2001).
The comparison between stellar masses based on evolutionary 
predictions (Color-Magnitude diagrams, CMDs) and pulsation predictions 
(Period-Mass-Radius relations) indicate that the discrepancy ranges 
from 10 to 20\% (Beaulieu et al. 2001; Bono et al. 2001; Bono, 
Castellani, \& Marconi 2002; Keller, \& Wood 2002). A similar discrepancy 
also applies to dynamical masses of binary Cepheids (Evans, Vinko, 
Wahlgren 2000; Petterson, Cottrell, \& Albrow 2004), but we still 
lack a homogeneous analysis of dynamical, evolutionary, and 
pulsational masses. At present, it 
is not clear whether such a discrepancy is due to a limit 
in the physical assumptions adopted to construct pulsation and 
evolutionary models or it is intrinsic. The latter working hypothesis 
is supported by the evidence that evolutionary models for 
intermediate-mass stars are typically computed by neglecting 
the mass loss during helium burning phases. This means that 
evolutionary models provide the mass which Cepheids have had along 
the Main Sequence, while pulsation models provide the actual mass 
of Cepheids. According to this undisputable fact, Caputo et al. (2004) 
performed a detailed analysis of well-observed Galactic Cepheids, and 
found that the discrepancy between evolutionary and pulsation masses 
depends on the pulsation period. In particular, the discrepancy 
increases when moving from long to short-period Cepheids. This 
finding might appear contrary to qualitative expectations, since 
long-period Cepheids are systematically redder than those with 
short-periods. However, the predicted behavior can be easily explained 
if we realize that short-period Cepheids spend a longer 
evolutionary time before and inside the instability strip (Brocato 
et al. 2004). This prediction, once supported by spectroscopic data,  
might have a substantial impact not only on predicted PL/PLC relations 
but also on the vexing question concerning the size of the core 
among intermediate mass stars.     

In order to constrain more quantitatively the impact of GAIA on the 
Cepheid distance scale, we performed a series of simulations using the 
Pisa Galactic model (Castellani et al. 2002; Cignoni et al. 2003). In 
particular, for the thin disk we assumed an exponential spatial distribution 
with a scale height of 250 pc (Mendez \& Guzman 1998). The normalization to the solar position 
relies on HIPPARCOS data (Cignoni et al. 2003). We adopted a Salpeter 
Initial Mass Function (IMF, $M^{-s}$ with s= 2.3) and the Star 
Formation Rate (SFR) predicted by chemical evolution models by Valle et al. (2004). 
Figures 1,2 and 3 show predicted CMDs for three arbitrarily selected regions 
covering an area of 1 square degree.  

%% Sample figure environment
\begin{figure}[!ht]
\begin{center}
\leavevmode
\centerline{\epsfig{file=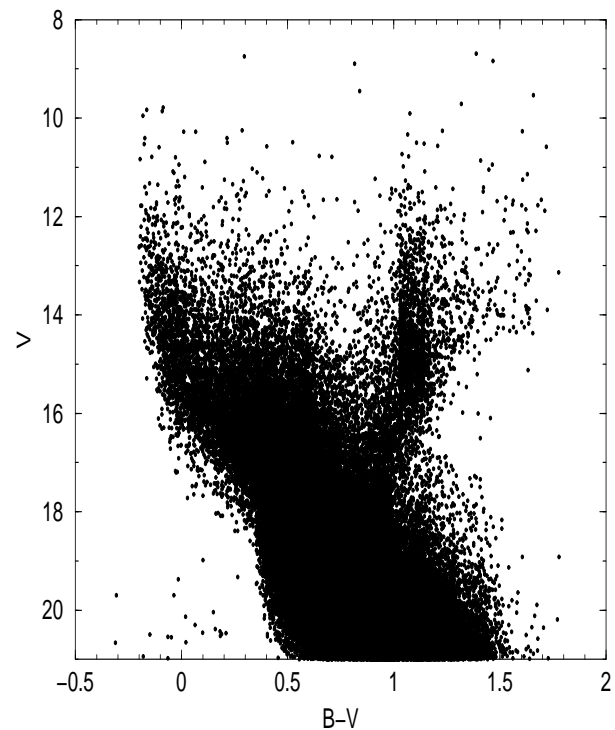, width=\linewidth, height=9.9cm}}
\end{center}
\caption{Simulated Color-Magnitude diagram for an arbitrary selected 
thin disk field located at Galactic longitude l=0 and latitude b=5. 
The field-of-view covered by simulations is 1 square degree.}
\label{fig.1}
\end{figure}

\begin{figure}[!ht]
\begin{center}
\leavevmode
\centerline{\epsfig{file=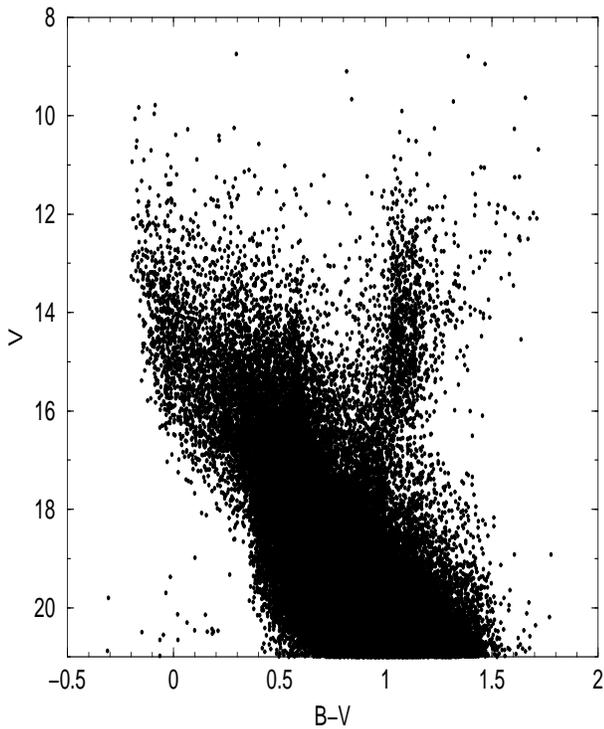, width=\linewidth, height=9.9 cm}}
\end{center}
\caption{Same as Fig. 1, but for l=20 and b=5.}
\label{}
\end{figure}

Note that current simulations have been performed by neglecting 
interstellar reddening and photometric errors on $B,V$ bands.
Moreover, we only included stars brighter than $V\sim 21$, since,  
according to current estimates (Jordi 2004, this volume) this is the 
limiting magnitude of GAIA photometry. In passing, we note that current 
simulations show three well-defined samples of disk White Dwarfs (WDs).   
($19 \le B$, $-0.4 \le B-V \le 0.2$). Note that such a number is only 
a lower limit to the number of WDs that GAIA will detect, since we 
only included CO-core WDs with H (DA) atmospheres (Castellani et al. 2002).       
We plan to include CO-core WDs with He (non-DA) atmospheres, and 
He-core WDs with a H atmosphere in the near future (Hansen \& Liebert 2003; 
Hansen 2004).    
To pinpoint classical Cepheids in the CMD, we adopted the analytical  
relations for the edges of the instability strip at solar chemical 
composition provided by Bono et al. (2004). The typical mean $B-V$ 
colors for Galactic Cepheids range from 0.3 to 1.1. Predictions have 
been transformed into the observational plane using the atmosphere 
models provided by Castelli, Gratton, \& Kurucz (1997).  
Current predictions based on Galactic models have already reached 
a mature phase (Robin et al. 2003; Bertelli et al. 1999). However, 
the star counts for relatively short evolutionary phases, such as 
the so-called blue loops of classical Cepheids, might depend on
the adopted local normalization. Therefore, we decided to normalize 
the expected number of Cepheids to the predicted number of MS stars 
with a stellar mass equal to 3.5 $M_\odot$ ($M_V=-0.3$ mag). Current 
simulations for the three selcted field predict 2000, 1500, and 100 
MS stars in the magnitude interval $M_V=-0.3\pm0.5$ and that the 
classical Cepheids correspond approximately to 2\% of this number, i.e. from 
2 to 40 per field. As a preliminary, conservative estimate, we assume that 
GAIA will observe a few thousand Galactic Cepheids. Short-period Cepheids 
have absolute visual magnitude of the order of $-2 \div -2.5$; therefore,  
GAIA will supply complete information for a large fraction of them. Note 
that, according to theoretical predictions based on nonlinear convective 
models (Fiorentino et al. 2002), the mean difference between SMC (mean 
metallicity Z=0.004) and Galactic (mean metallicity Z=0.02) Cepheids 
is roughly equal to 0.4 mag at $\log P =1$. A series of random extractions 
based on the OGLE catalogue for Magellanic Cepheids (Udalski et al. 2001) 
indicates that for a sample of 1500 Cepheids covering a metallicity range of 
1 dex (Z=0.002 - 0.02) and for which are available: {\em i)} geometric 
distances with an accuracy better than 2\% ($\sim 0.04$ in distance
modulus); {\em ii)} metal abundances with an accuracy better than 0.2 dex;
{\em iii)} reddening estimates with an accuracy better than 0.02 mag, 
will allow us not only to constrain the metallicity dependence in the optical
bands with an accuracy better than a few hundreths of magnitude, but also to 
constrain the fine structure of both PL and PLC relations over the entire 
period range. 

\begin{figure}[!ht]
\begin{center}
\leavevmode
\centerline{\epsfig{file=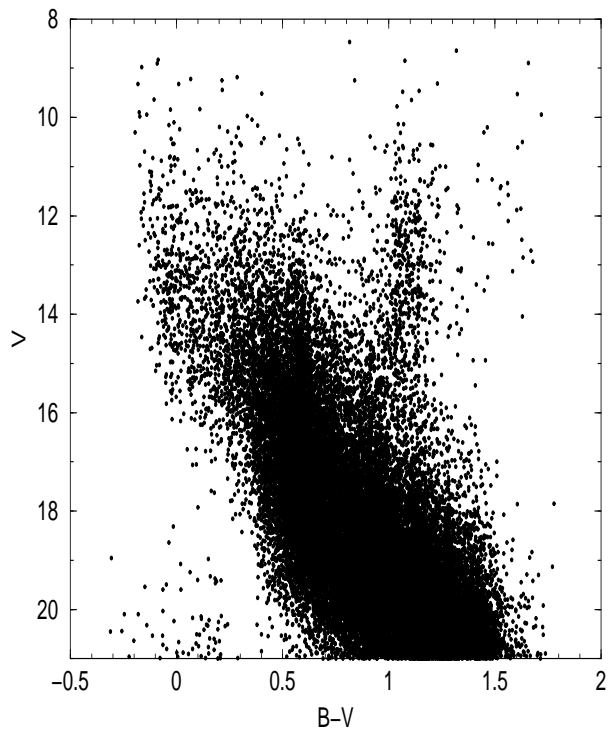, width=\linewidth, height=9.9 cm}}
\end{center}
\caption{Same as Fig. 1, but for l=180 and b=5.}
\label{}
\end{figure}

Classical Cepheids are also excellent tracers for intermediate-age stellar 
populations, and are widely used to estimate the metallicity
gradient across the Galactic disk (Bono 2003a,b) and to constrain Galactic 
rotation (Pont et al. 1997; Metzger et al.  1998). Moreover, they can also 
be adopted to estimate the possible occurrence of age gradients. Dating back 
to Kippenhahn \& Smith (1969) and to Meyer-Hofmeister (1969), it has been  
recognized that an increase in the pulsation period implies an increase in 
the stellar mass, and therefore a decrease in the Cepheid age. On the basis 
of these arguments, several Period-Age (PA) relations have been derived  
(Efremov 1978; Tsvetkov 1989; Magnier et al. 1997). Stellar ages based 
on PA relations present two substantial advantages when compared with the 
isochrone fitting method:\\  
\begin{figure}[!ht]
\begin{center}
\leavevmode
\centerline{\epsfig{file=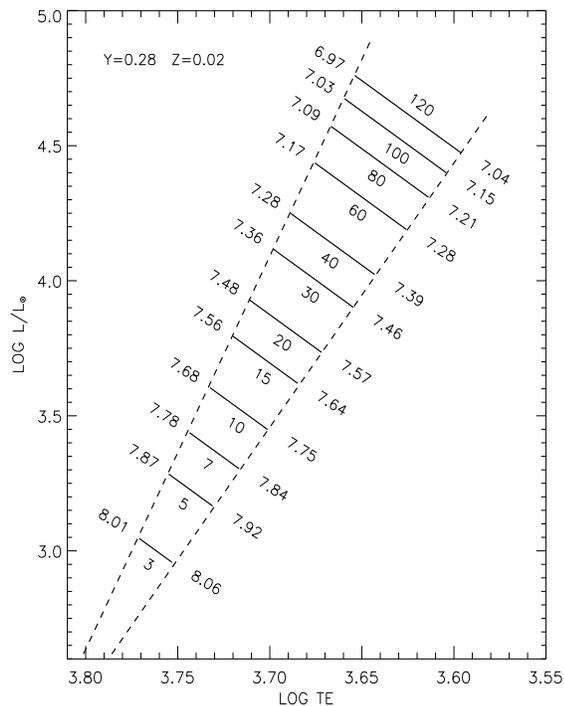, width=\linewidth, height=9.9 cm}}
\end{center}
\caption{Predicted instability strip for fundamental Cepheids at 
solar chemical composition. Logarithmic Cepheid ages along the blue (hot) 
and the red (cool) edge of the instability strip are labeled. Solid lines 
show iso-period lines.}
\label{}
\end{figure}
{\em i)} the period depends on neither the distance modulus, nor the 
cluster reddening, nor the photometric calibration; 
{\em ii)} the PA relation can be applied to individual objects, and therefore 
relative ages and age gradients can be estimated with a high spatial resolution.\\    
On the other hand, the main drawback is that Cepheid ages based on the
PA relations rely on the assumption that the Cepheid instability strip 
has a negligible width in color. This assumption is plausible 
in the short-period range, but the ages of long-period ($\log P > 1$) 
Cepheids should be estimated on the basis of a Period-Age-Color (PAC) 
relation. Figure 4 shows the predicted instability strip for fundamental 
Cepheids with solar chemical composition (Z=0.02). Cepheid ages are labeled 
along the blue and the red edge (dashed lines), while solid lines show 
iso-period lines. Data plotted in this figure show that the difference is 
smaller than 10\% for periods shorter than $ P < 3$ days and larger than 15\% 
for periods longer than $ P > 120$ days. A new spin to this intersting approach 
has been recently provided by Bono et al. (2004), who estimated new theoretical 
PA and PAC relations for both fundamental and first overtone Cepheids. They 
applied the new relations to large samples of Galactic and Magellanic cluster 
Cepheids, and found that the difference between cluster ages based on 
isochrones and mean ages based on PA and PAC relation is smaller than 20\%.

%%%%%%%%%%%%%%%%%%%%%%%%%%%%%%%%%%%%%%%%%%%%%%%%%%%%%%%%%%%%%%%%%
\section{RR Lyrae stars}

Horizontal Branch stars and in particular RR Lyrae stars are very good 
tracers of old, low-mass stars (Suntzeff et al. 1994; Kinman et al. 1996; 
Wilkinson \& Evans 1999; Bragaglia et al. 2004, this volume). The main advantage,  
when compared with classical Cepheids, is that they are ubiquitous in the 
Galactic spheroid (halo, thick disk, bulge), and are present both in early and 
late type galaxies (van den Bergh 2000). However, they are at least a couple 
of magnitude fainter than classical Cepheids. Therefore, at fixed limiting 
magnitude, Cepheids allow us to cover a sky volume which is at least 90\% 
larger. Notwithstanding this, distances based on RR Lyrae are fundamental 
for constraining the occurrence of systematic errors as well as for shedding new 
light on the input physics adopted for evolutionary and pulsation models 
of low-mass stars (Wilkinson et al. 2004).    

In order to constrain more quantitatively the impact of GAIA on the 
RR Lyrae distance scale, we performed a series of simulations using the 
same Galactic models adopted for Cepheids.  In particular, 
for the thick disk we assumed an exponential spatial distribution 
with a scale height of 900 pc (Gilmore \& Reid 1983; Santiago et
al. 1996). 
% Sample figure environment
\begin{figure}[!ht]
\begin{center}
\leavevmode
\centerline{\epsfig{file=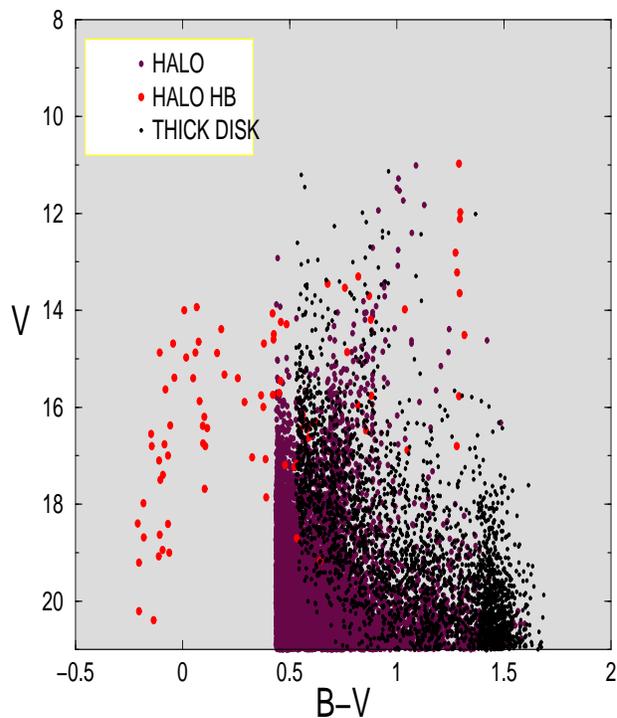, width=\linewidth, height=9.9 cm}}
\end{center}
\caption{Simulated Color-Magnitude diagram for an arbitrary selected 
halo plus thick disk field located at Galactic longitude l=0 and latitude b=60. 
The field-of-view covered by simulations is 5 square degrees. Thick circles 
mark the positon of HB stars.}
\label{fig.1}
\end{figure}
For this component, we assumed a constant SFR between 8 and 10 Gyr and a mean 
metallicity Z=0.006 (Gilmore et al. 1995). The normalization 
at the solar position was assumed equal to the 3\% of the thin disk. For the 
halo, we assumed a density distribution {\em a la} de Vacouleurs, with a half-light 
radius of 3 Kpc. According to Robin et al. (2000), we assumed a 
normalization at the solar position of $1.6\times 10^{-4}$ stars per parsec 
cubic. Moreover, we assumed a typical age of 12 Gyr and a mean metallicity 
of Z=0.001. Note that, to populate the HB, we also assumed a mean mass of 
0.65 $M_\odot$ and a spread in mass of 0.02 $M_\odot$. The reader interested 
in a detailed discussion concerning the computation of synthetic HB diagrams 
is referred to Catelan et al. (2004) and Cassisi et al. (2004, and references 
therein). To pinpoint RR Lyrae stars in synthetic CMDs, we adopted pulsation 
predictions provided by Bono et al. (2003). Once again, we neglected interstellar 
reddening and intrinsic photometric errors.    

Figures 5,6, and 7 show simulations along three arbitrary selected halo 
fields. To avoid deceptive uncertainties due to the local normalizations,  
we estimated the number of both HB and RR Lyrae stars as the ratio between 
HB stars and the number of MS stars with $M=0.78 M_\odot$ ($M_V=4.1\pm0.5$ mag).  
In particular, we found that the MS stars across this magnitude interval are 1354,
491, and 286 in the three simulations, while the number of HB stars is $\approx 10$\% 
of the entire sample. Roughly the 14\% of HB stars are RR Lyrae stars. The typical 
mean $B-V$ colors of RR Lyrae stars range from $\sim 0.2$ (first overtone pulsators) 
to $\sim 0.5$ (metal-rich fundamental pulsators). It is worth noticing that figures 5,6, 
and 7 also show samples of hot ($B-V \le 0.1$) HB stars. On the basis of detailed 
spectroscopic measurements, Preston, \&  Sneden (2000) draw the attention on a sample 
of sixty-two blue metal-poor stars. Interestingly enough, more than 60\% of these stars are 
binaries and 50\% of them appear to be Blue Stragglers. It is plausible to assume 
that the some of the remaining binaries could be hot HB stars. The detection of these 
objects in binary systems is very promising, since we lack accurate measurements of 
dynamical masses for HB stars. However, we emphasize the fact that current simulations 
are very preliminary, since they rely on crude approximations. Moreover, the number of RR Lyrae 
strongly depends on the assumptions adopted when constructing synthetic HB diagrams. 
Current uncertainties are mainly due to the fact that we lack an exhaustive knowledge 
of the astrophysical parameter(s) which, together with the metallicity,  govern 
HB morphology, i.e. the so-called second parameter problem (Sandage 1993a,b,c; 
Richer et al. 1996; Buonanno et al. 1998; Castellani et al. 2003; Castellani et al. 2004). 
Moreover and even more importantly, we lack empirical and theoretical insights 
on the efficiency of mass-loss along the RGB, as well as on its dependence on metal 
abundance.   

\begin{figure}[!ht]
\begin{center}
\leavevmode
\centerline{\epsfig{file=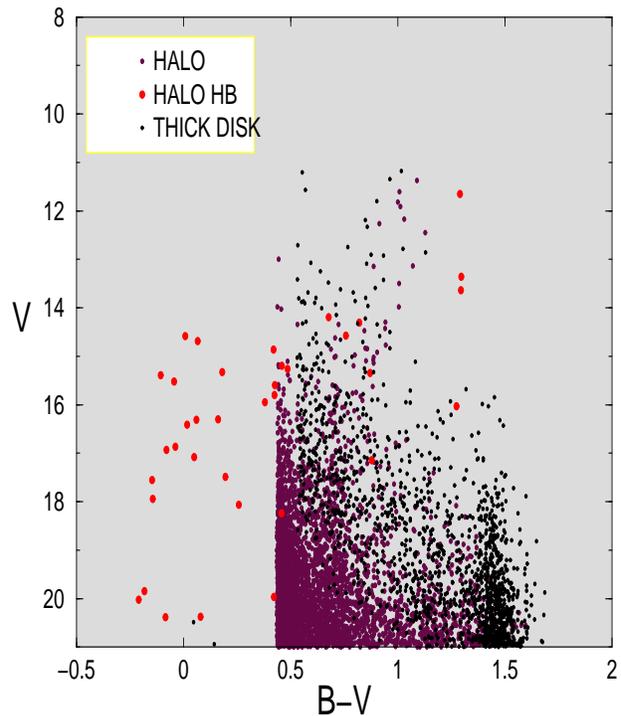, width=\linewidth, height=9.9 cm}}
\end{center}
\caption{Same as Fig. 5, but for l=0 and b=90.}
\label{}
\end{figure}

Preliminary estimates suggest that a sample of $\approx 1000$ RR Lyrae covering a 
metallicity range of 2 dex (Z=0.0002 - 0.02) and for which are available: 
{\em i)} geometric distances with an accuracy better than 2\% ($\sim 0.04$ in 
distance modulus); {\em ii)} metal abundances with an accuracy better than 0.2 dex;
{\em iii)} reddening estimates with an accuracy better than 0.02 mag, 
will allow us to assess on a quantitative basis the dependence of RR Lyrae 
stars on metallicity and to estimate the evolutionary effects. This means that 
we can calibrate all the methods currently adopted to estimate RR Lyrae distances 
(Baade-Wesselink, statistical parallaxes, K-band PL relation, First Overtone Blue Edge). 
On the other hand, HB stars will supply fundamental constraints on the input physics adopted 
to construct HB models (Castellani, \& degl'Innocenti 1993; Wilkinson et al. 2004). In particular, 
the ratio between HB and Red Giant Branch stars, the so-called R parameter, will supply 
firm constraints on the cross-section of the fundamental $^{12}C\,(\alpha, \gamma)\,^{16}O$ 
reaction rate (Salaris et al. 2004, and references therein) and on the efficiency 
of mixing processes during central helium-burning phases (Caputo et al. 1989; 
Straniero et al. 2003).    

\begin{figure}[!ht]
\begin{center}
\leavevmode
\centerline{\epsfig{file=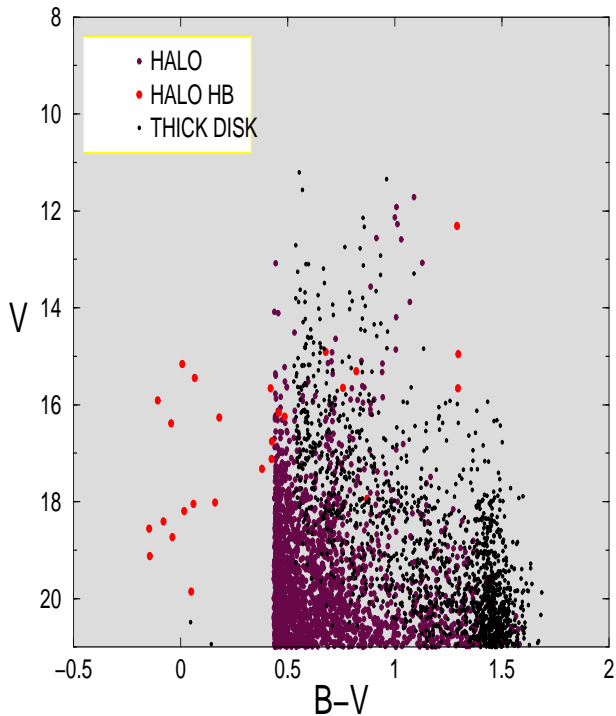, width=\linewidth, height=9.9 cm}}
\end{center}
\caption{Same as Fig. 5, but for l=180 and b=60.}
\label{}
\end{figure}

The $R$ parameter is currently adopted to provide an empirical 
upper limit to the helium abundance (Zoccali et al. 2000; 
Salaris et al. 2004). However, robust empirical arguments 
(Sandquist 2000) suggest that the $A$ parameter introduced by 
Caputo et al. (1983) might be a very accurate helium indicator.  
The $A$ parameter is related to the ML relation of RR Lyrae 
stars, i.e.  $A=\log (L/L_\odot)-0.81 \log (M/M_\odot)$, and 
relies on theoretical evidence according to which an increase in helium 
abundance causes an increase in the luminosity and in the mean 
mass of RR Lyrae stars. Empirical estimates of this parameter 
can be obtained using the pulsation period and the mean effective 
temperature of RR Lyrae stars together with the pulsation relation 
(van Albada \& Baker 1973). Unfortunately, current empirical and 
theoretical color-temperature relations are still affected by systematic 
uncertainties (Silbermann, \& Smith 1995; Cacciari et al. 2000) which 
have severely limited the use of the $A$ parameter. Homogeneous 
photometric and spectroscopic data will be collected by GAIA,  
and will allow us to calibrate these relations and to supply 
accurate estimate of helium abundance in Local Group stellar 
systems.

%%%%%%%%%%%%%%%%%%%%%%%%%%%%%%%%%%%%%%%%%%%%%%%%%%%%%%%%%%%%%%%%%%%%%%%%%%
\section{Final Remarks}

We have already discussed the crucial role that GAIA will 
play concerning the distance scale and stellar populations. 
However, we have to proceed cautiously. Accurate measurements of stellar 
astrophysical parameters and distances require a detailed  
mapping of instellar extinction across the Galactic spheroid. 
New spectroscopic approaches have been recently suggested 
(Katz et al. 2004), but we also need to develop new methods
that rely on reddening free indices and on luminosity amplitude and 
mean colors of RR Lyrae stars (Kovacs, \& Walker 2001; Piersimoni et al. 2002). 
This requirement is mandatory, since GAIA photometry will allow us to measure 
stars which are roughly two order of magnitude fainter than spectroscopy. 
The use of both broad and intermediate-band photometry together with a very 
accurate and stable photometric absolute zero-point calibrations might 
be the keystone  for this crucial issue.   

During the last few years, evolutionary and pulsation predictions 
are shaking off their speculative nature thanks to thorough comparisons 
between theory and observations. The mosaic CCD cameras and the 
multi-fiber spectrographs available at the 8m class telescopes will 
certainly improve the empirical scenario for stellar systems in the 
Local Group. Obviously, GAIA will play a fundamental role for an 
understanding of the formation and evolution of low-density regions of 
the Galactic spheroid (halo, thin and thick disk). The same outcome 
applies to the calibration of distance indicators and stellar tracers.  
These are two fundamental steps in view of a quantitative understanding 
of stellar populations in galaxies beyond the Local Group. In this 
investigation, we did not mention stellar populations in the Galactic bulge. 
No doubt they are the key to understand stellar structures in the 
metal-rich ($Z \ge 0.02$) regime, i.e. the analog of stellar populations 
in elliptical galaxies. The only other stellar system in the Local Group 
which can help us to understand these stellar populations is M32. 

Detailed estimates concerning the intrinsic accuracy of current distance 
determinations and predictions concerning the open problems which we shall  
debate in ten years are very difficult. During the writing of this 
manuscript, we remembered an interesting paper by Lynden-Bell (1972) focused 
on Galactic distance determinations. We quote a few sentences:\\ 

{\em ... I must prophesy the future of astronomical distance measurements,
but first I must introduce you to two unwanted, uninvited guests. Dr Realist 
estimates errors not by what observers state, but by changes between values 
derived by different people. Dr Pessimist thinks of errors due to inadequate 
concepts as well as taking a less rosy picture of the errors than Dr Realist.}         

Following the Dr Realist approach we should support the view that during the 
next few years we should be able to provide absolute distances with an accuracy 
better than 5\% using both Cepheids and RR Lyrae stars. On the other hand, 
according to the Dr Pessimist approach this might be a deceptive prediction. 
What is very promising in this context is that GAIA will supply geometrical 
distances with meaningful individual error bars, it will thus provide the 
chance for a historical convergence between the approaches of both 
Dr Realist and Dr Pessimist.

\section*{Acknowledgments}
It is a pleasure to thank F. Caputo, V. Castellani, and S. Shore
for a detailed reading of an early draft of this manuscript. We also 
wish to acknowledge the RVS working group and the Variable Stars 
working group for many useful discussions and suggestions concerning 
the topics of this paper. 
This work was partially supported by MIUR/PRIN~2003 in the framework 
of the project: ``Continuity and Discontinuity in the Galaxy Formation`` 
and by INAF/PRIN~2003 in the framework of the project: ``The Large 
Magellanic Cloud: a laboratory for stellar astrophysics``.

\end{document}